\documentstyle[aps,prl,multicol]{revtex}
\input{epsf}
\begin{document}
\draft \title{Drifting Pattern Domains in a Reaction-Diffusion System 
with Nonlocal Coupling}

\author{  Ernesto M. Nicola,
Michal Or-Guil, Wilfried Wolf and Markus B\"{a}r}

\address{ Max-Planck-Institut f\"ur Physik komplexer Systeme, \\
        N\"othnitzer Stra{\ss}e 38, D-01187 Dresden, Germany  } 
\date{May 19, 2000}
\maketitle

\begin{abstract}

Drifting pattern domains (DPDs), {\it i.e.} moving
localized patches of traveling 
waves embedded in a stationary (Turing) pattern 
background and {\it vice versa}, are observed in simulations of a
reaction-diffusion model with nonlocal coupling. 
Within this model, a region of bistability between Turing patterns and 
traveling waves arises from a codimension-2 Turing-wave bifurcation
(TWB).
DPDs are found within that region in a substantial distance from the TWB.
We investigated the dynamics of 
single interfaces between Turing and wave patterns.
It is found that DPDs exist due to a locking of the interface velocities,
which is imposed by the absence of space-time defects near these interfaces.  
\end{abstract}


\begin{multicols}{2}



{\em - Introduction - }
Pattern forming processes in nonequilibrium systems 
can be classified according to the primary instability of the
spatially homogeneous state. 
Ref. \cite{Cross-Hohenberg} distinguishes three basic types of 
instabilities in unbounded systems: 
{\it (i)} spatially periodic and stationary in time, 
{\it (ii)} spatially periodic and oscillatory in time and {\it (iii)} 
spatially homogeneous and oscillatory in time. 
Within the reaction-diffusion literature, these instabilities are known
as Turing, wave and Hopf bifurcation, respectively. 

Many chemical and biological patterns are well captured by so called
activator-inhibitor models \cite{Mikhailov} describing the 
dynamics of two reacting and diffusing substances with two coupled
partial differential equations. 
In such two component reaction-diffusion models only Turing and Hopf 
instabilities are possible.  
Recently, numerical investigations of chemical
reaction-diffusion systems with three components \cite{Zhabotinsky} 
and nonlocal coupling \cite{MexPRL} have yielded the occurrence of wave
instabilities and the corresponding patterns. 
A universal description of patterns near these instabilities
is achieved within the framework of amplitude equations 
\cite{Cross-Hohenberg,MvHecke}. 

Here, we study a simple FitzHugh-Nagumo model with  
inhibitory nonlocal coupling that is obtained as a limiting case
of a three component reaction-diffusion system.
It describes the interaction of an activator species with 
an inhibitor.
For slow inhibitor diffusion (compared to the activator diffusion), 
the model exhibits wave instabilities, 
while, for fast inhibitor diffusion, Turing instabilities are found.
The two instabilities occur simultaneously  at a codimension-2 Turing-wave
bifurcation (TWB). 
Such a situation has been found earlier within a model for
binary convection \cite{Schoepf} and is a generalization of the well
investigated Turing-Hopf instability in reaction-diffusion systems 
\cite{TuHo}. 
Basic properties of a  TWB  have  been studied  
theoretically in amplitude equations \cite{Walgraef} as well as 
experimentally in a one-dimensional gas-discharge system \cite{Purwins}. 
In our model, we find a pattern previously unknown in
\begin{figure}[t]
\label{FIG 1}
\epsfxsize=80mm
\centerline{\epsffile{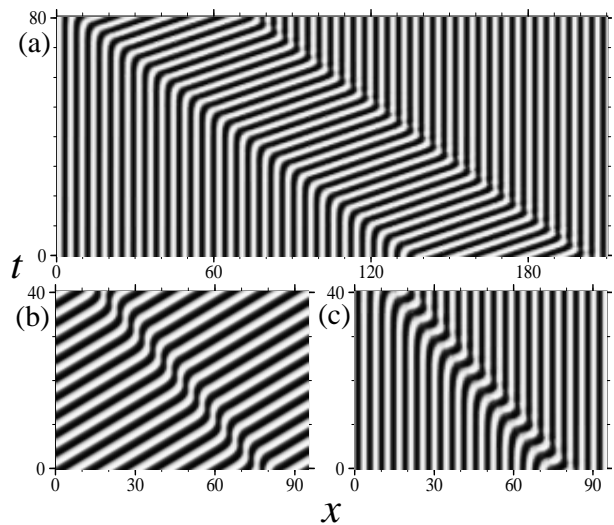}}
\vspace{2mm}
\caption[]{Space-time plots of the field $u$ in grey scale for three examples of 
DPDs found in numerical simulations of Eqs.\ (1). 
In all three cases $a=6.0$, and only a part of the system of length $L=409.6$
is shown. (a) Large DPD with $\delta=0.84$. (b),(c) DPDs consisting of a
single cell of Turing and wave respectively. In (b) $\delta=0.80$ and in (c)
$\delta=0.91$. Other parameters, see \cite{codim2}.
}
\end{figure}
\noindent reaction-diffusion systems: drifting pattern domains (DPDs), 
{\it i.e.} 
localized patches of traveling waves embedded in a
 Turing background and {\it vice versa} (see Fig. 1). 
These patches have
constant width and move (drift) with constant speed. 
As they drift along, maxima of
the concentrations of activator and inhibitor are conserved 
beyond both  boundaries of the DPD,
where Turing and wave patterns are joined together.
Consequently, formation of space-time defects by coalescence of maxima and
minima is prevented.
Similar patterns  have been reported in a variety of hydrodynamical 
experimental systems, see {\it e.g.}  \cite{Fless,LiqCol} and have been related
to secondary instabilities  (parity breaking) 
of stationary patterns \cite{Fless,GoldPB}.
DPDs exist in a broad region of the para-
\begin{figure}
\label{FIG 2}
\epsfxsize=80mm
\centerline{\epsffile{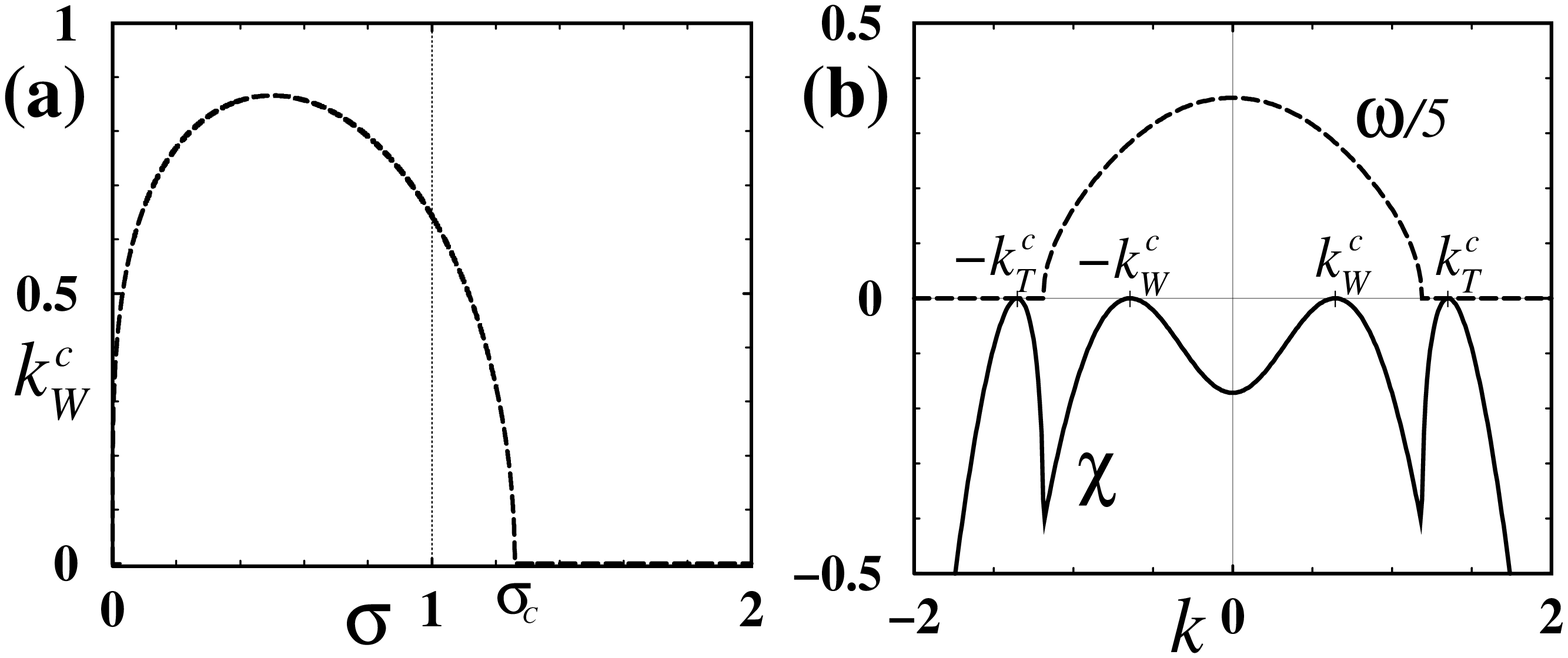}}
\caption[]{
(a) Critical wavenumber $k_W^c$ against inverse nonlocal coupling range 
$\sigma$. (b) Real ($\chi$) and imaginary part ($\omega$)  of $\lambda(k)$ at the
TWB, for parameters see \cite{codim2}.}
\end{figure}
\noindent meter space, but appear only in a substantial distance to the onset of pattern formation. 
Their existence region is characterized by 
bistability between waves and Turing patterns. 
Near the boundary of DPD existence small DPDs containing only a 
single wave or Turing ,,cell'' inside a background of the respective other state are encountered 
(see Figs. 1b,c). 
Outside the DPD existence region,
 the width of pattern domains shrinks or expands; 
these transient domains exhibit defects at the interfaces. 
Large DPDs  are composed of two interfaces 
separating wave and Turing patterns.
In this Letter we will study the dynamics of such interfaces.
These interfaces typically select the wavenumber of the invading domain and
form one parameter families characterized by the wavenumber 
of the invaded domain. 
Far away from the TWB, interfaces can exhibit a locking mechanism
of their velocities due to the absence of defects. 
This locking implies that the interface velocity is fixed by 
the wavenumbers and frequency of the patterns that they separate. 
If both interfaces in a DPD are locked, 
they have to travel with equal speed. 
This mechanism is responsible for the existence of DPDs with constant 
width.

{\em - Model equations and linear stability  - }
We study a variant of the FitzHugh-Nagumo equation supplemented by an 
inhibitory nonlocal coupling in the dynamics of the activator $u$
\begin{eqnarray}
\partial_{t}u & = & a u +\beta  u^2 - \alpha u^3 -bv  
+ \partial_{x}^2 u\nonumber  \\ 
& &- \mu \int_{-\infty}^{+\infty} e^{-\sigma |x-x^{'}|}
u(x^{'},t)dx^{'} 
\nonumber \\
\partial_{t}v &=& c u - d v  + \delta  \partial_{x}^2 v. \label{eq_nlc}
\end{eqnarray}
Eqs. (1) represent a limiting case of a three variable model involving the  
activator $u$, the inhibitor $v$ and an additional fast inhibitor \cite{3var}. 
Related three variable models  have been introduced previously to describe 
pattern formation on sea shells and in cell biology  \cite{Meinhardt} as well 
as spot dynamics in gas discharges \cite{Discharge} and concentration patterns
in heterogeneous catalysis \cite{Moshe}. 
Here, the emphasis is on the onset of pattern formation resulting
from destabilization of a single homogeneous steady state. 
Eqs. (1) possess the trivial homogeneous fixed point 
${\bf u}_0 {\stackrel{{\scriptscriptstyle \rm def}}{=}} (u_0,v_0)^T = (0,0)^T$ 
for all parameter values. 
Here, we consider the regime where this fixed point is the only one present,
{\it i.e.} $a < bc/d + 2 \mu/\sigma$ and consider perturbations 
$\propto e^{ikx - \lambda(k) t}$, where $\lambda(k) = \chi(k) + i \omega(k)$. 
The growth rates $\lambda(k)$ are given by the eigenvalues of the Jacobian.
Linear stability analysis reveals that Eqs. (1) exhibit wave
instabilities if the nonlocal coupling is of sufficiently 
long range $\sigma <  \sigma_c = (2 \mu / (1 + \delta))^{1/3}$. 

In the following we vary the control parameters $a$ and $\delta$; the
,,driving force'' $a$ represents the kinetics, whereas the ratio of diffusion 
coefficients $\delta$ describes the spatial coupling in the medium.  
All other parameters of Eqs. (1) have been  fixed \cite{codim2}.
For the wave bifurcation, 
the critical wavenumber $k_W^c$ and parameters $a_W, \delta_W$ are 
obtained from the condition $\lambda (k_W^c) = \pm i \omega_0$ where the perturbation with $k_W^c$ is the fastest growing mode with 
$ (k_W^{c})^2 = \sqrt{\frac{2\mu\sigma}{1+\delta}}  -{\sigma^2} $.
Note, that for both $\sigma = \sigma_C$ and $\sigma = 0$ 
(global coupling limit) the critical wavenumber is $k_W^c = 0$, see Fig. 2a. 
Similarly, a competing Turing instability appears for a critical
parameter $a_T$ with a wavenumber $k_T^c$, where the leading eigenvalue 
$\lambda(k_T^c) = 0$.
For large enough driving $a$,
the wave instability appears for small $\delta$, while for
large $\delta$ the Turing instability destabilizes the homogeneous 
state. 
For the chosen parameter values, the system exhibits a TWB point (see Fig. 3a and \cite{codim2}).
For the corresponding $\lambda(k)$, see Fig. 2b.

{\em - Weakly nonlinear analysis - }
Near the TWB,  we can write ${\bf{u}}{\stackrel{{\scriptscriptstyle \rm def}}{=}}(u,v)^T $ as a perturbative
expansion around $\bf{u}_0$ using a small parameter $\varepsilon$, indicating the distance to
the instability threshold: ${\bf{u}}={\bf u}_0 + \varepsilon
{\bf{u}}_1+ \varepsilon^2 {\bf{u}}_2+\varepsilon^3{\bf{u}}_3 +\cdots$ 
and use the following multiple scale ansatz: 
\cite{TwoTimes}
\begin{eqnarray*}
{\bf u}_1\!&=&\! [A(\!X\!,\!T_{_1}\!,\!T_{_2}\!){\bf U}_{\!\!A}
e^{i(\omega_0 t+k_W^c x)}+ B(\!X\!,\!T_{_1}\!,\!T_{_2}\!)
{\bf U}_{\!\!B}e^{i(\omega_0 t-k_W^c x)}  \nonumber \\ 
& &+{\cal R}(\!X\!,\!T_{_1}\!,\!T_{_2}\!){{\bf U}_{\!\!{\cal R}}}e^{ik_T^c x} +
c.c.]/2. \nonumber 
\end{eqnarray*}
This leads to a set of coupled equations for the amplitudes $A$, $B$ and
$\cal{R}$ for left-, right-going waves and Turing pattern that
depend on slow time and space variables. 
After reestablishing the original time
and space variables and performing further $\varepsilon$-independent
scaling, one obtains:
\begin{eqnarray}
 \partial_{t}{\cal R}&=&\eta {\cal R}
 - |{\cal R}|^{2}{\cal
R}  + \xi \partial_{x}^2{\cal  R} - \zeta (|A|^{2}+ |B|^{2}){\cal
R}
\nonumber \\
\partial_{t}A+ c_g \partial_{x} A&=& \rho A+ (1+i c_1) \partial_{x}^2 A -
(1-ic_3)|A|^{2}A \nonumber \\
& &- g (1-ic_2)|B|^{2}A-\nu(1-i\kappa)|{\cal R}|^{2}A
\nonumber \\
\partial_{t}B- c_g \partial_{x} B&=& \rho B +(1+i c_1) \partial_{x}^2 B-
(1-ic_3)|B|^{2}B \nonumber \\
& &- g (1-ic_2)|A|^{2}B-\nu(1-i\kappa)|{\cal R}|^{2}B.
\end{eqnarray}
For the detailed values of all coefficients, see \cite{coeff}.
Note, that the nonlocal term of Eqs. (1) only enters into the 
diffusion coefficients of Eqs. (2) and does 
not give rise to a nonlocal term in Eqs. (2). 
Knowledge of  the coefficients of Eqs. (2) allows analytical
predictions of the pattern dynamics.
Here, traveling waves are always preferred over standing waves ($ g  >
1$, see \cite{Cross-Hohenberg}) and bistability between wave and 
Turing patterns is found ($\nu \zeta > 1$). 
In this bistability region in parameter space (see Fig.\ 3a), 
a family of stable Turing patterns and two 
families of stable left- and right-traveling waves parametrised by their 
corresponding wavenumbers coexist. 
To get further insight, we take a closer look at the dynamics of
single interfaces separating domains of Turing and wave patterns. 

{\em - Interface dynamics. -}
With suitable initial conditions, 
a moving interface between Turing and wave patterns will be formed in 
simulations of Eqs. (1).
We can distinguish two types of interfaces
depending on whether the phase velocity of the waves points towards the interface or
away from it. 
This classification is independent from the direction in which the interface is moving.
In the following, we will call the first type {\it inward-interfaces} and
the second {\it outward-interfaces}. 
Fig.\ 3b and 3c show examples of the latter type.

Near the TWB, we have studied general properties of such interfaces 
in amplitude equations (3) by counting arguments \cite{MvHecke}
as well as by direct numerical simulations. 
Counting arguments are applied to ordinary differential equations 
obtained from a coherent structure ansatz in a comoving
frame. 
We observe that, typically, the wavenumber of the invaded domain remains
constant, while it adapts in the invading domain.  
In other words, the interface selects a particular wavenumber for 
the invading state, while the initial wavenumber of the invaded 
state is a free parameter. 
The velocity of the interface is a function of this parameter. 
For Turing patterns the selected wavenumber is always the critical one, 
{\it i. e.} $k_T^{sel} = k_T^c$, while for waves typically 
$k_W^{sel} \ne k_W^c$ and therefore $\omega^{sel} \ne \omega_0$. 
This is valid for both inward- and outward-interfaces. 
Thus, we typically have two one-parameter families of interfaces for a
given point in parameter space. 
These results are confirmed by numerical simulations of the nonlocal model (1)
near the TWB. 

Simulations of interfaces in 
Eqs. (1) far away from the TWB, show qualitatively similar
behavior with respect to the selected wavenumbers. 
In addition,  
interfaces far away from the TWB may
exhibit a locking mechanism of their velocities, which are 
fixed by the wavenumbers and frequencies of the Turing and wave domains.
More specifically, the selected velocity is determined by the 
absence of defects at the interface.
For geometrical reasons an interface without defects, that 
connects a wave state with wavenumber $k_W$
and frequency $\omega$ and a Turing state with $k_T$, has a speed 
$|v_{lock}|= \omega / (k_T - k_W)$. 
This velocity locking mechanism is found for both types of interfaces. 
Two examples of areas where locking occurs  ({\it locking tongues}) 
in parameter space for inward- and outward-interfaces are given in Fig. 3a. 
A locked outward-interface is displayed in Fig. 3b. 
Outside the tongue the 
outward-interfaces display phase slips (see Fig. 3c). 
The area of the locking tongues depends only weakly on the front 
parameter $k_W$ for inward-interfaces and $k_T$ for outward-interfaces.
Note, that the locking tongues open at a sub-
\begin{figure}[t]
\label{FIG3}
\epsfxsize=70mm
\centerline{\epsffile{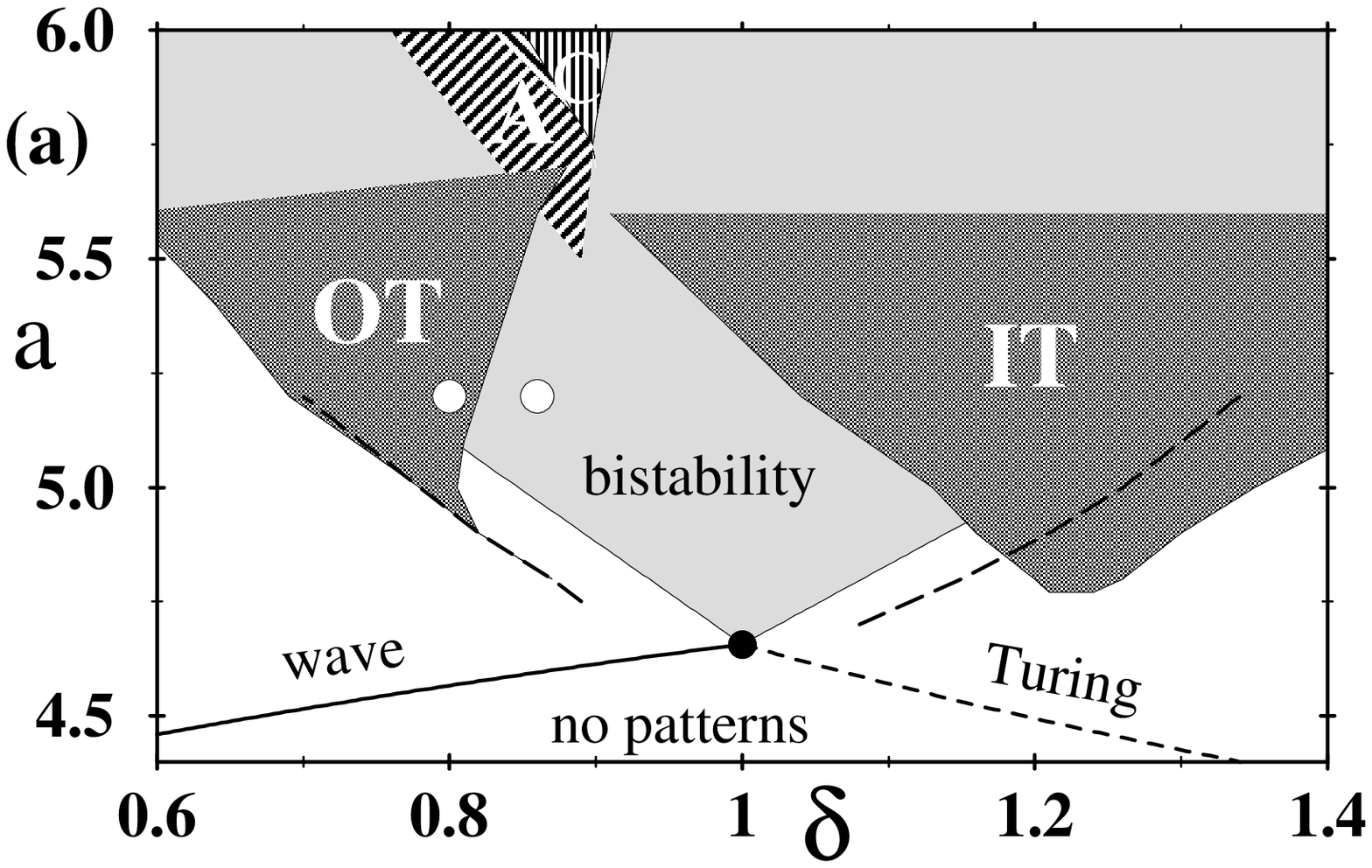}}
\vspace{-4mm}
\epsfxsize=85mm
\centerline{\epsffile{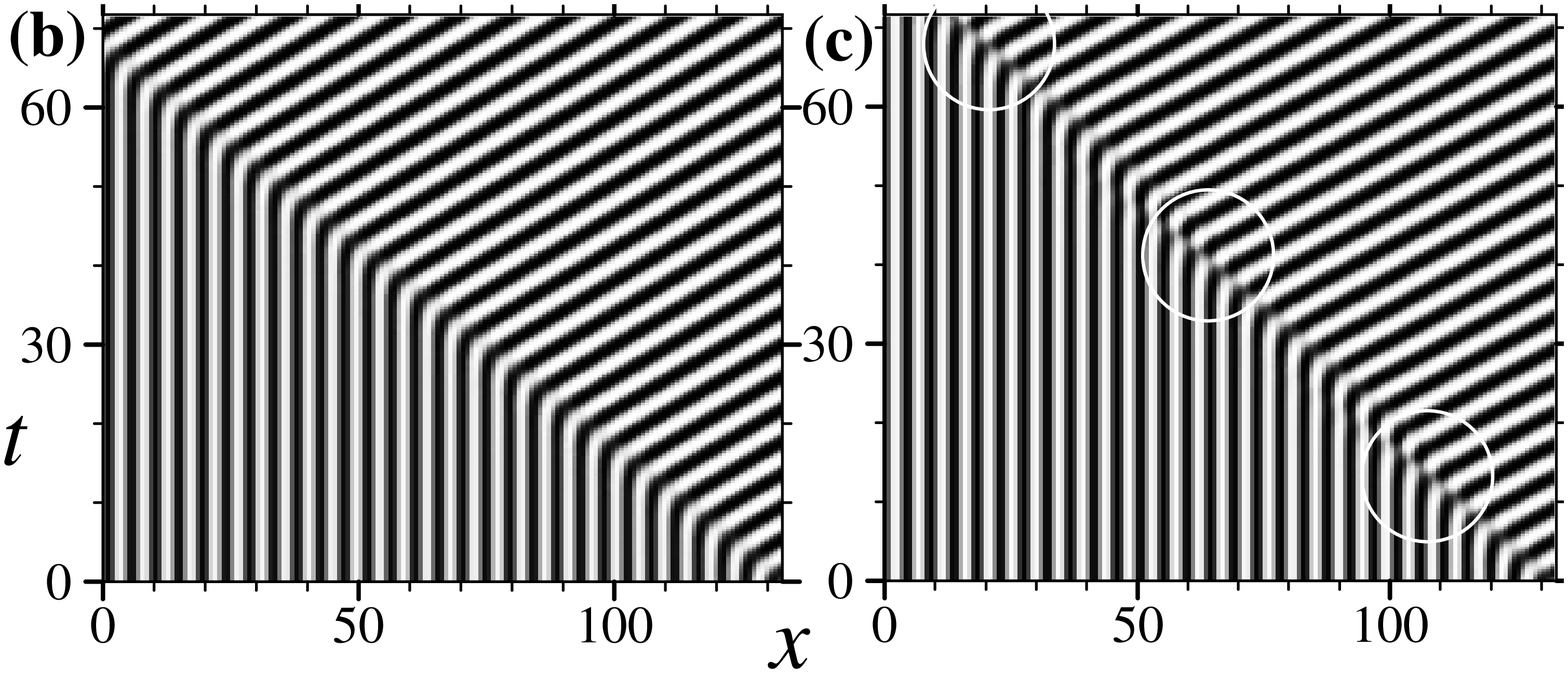}}
\vspace{1mm}
\caption{(a) Parameter space $a$-$\delta$ near the TWB point (black circle).  
The light grey region indicates bistability between the Turing and
wave pattern with $k_T^c, k_W^c$ as predicted from Eqs.\ (2). 
Dark grey regions correspond to two examples of locking tongues for an
outward-interface with $k_T=k_T^c$ (region {\bf OT}) and an 
inward-interface with $k_W= k_W^c$ (region {\bf IT}).
These two tongues are shown only up to $a\approx 5.6$.
The dashed lines show where the selected velocities of interfaces in 
Eqs. (2) coincide with $v_{lock}$. 
White circles indicate parameter values of simulations shown in (b) and (c). 
See Fig. 4 for a description of the dashed regions.
(b) and (c)  show space-time plots of $u$ in grey scale
from simulations of Eqs.\ (1) showing outward-interfaces for $a=5.2$. 
In (b) an example inside the locking tongue is shown for $\delta=0.80$ and in (c) an
interface outside the tongue exhibiting defects (inside the white circles) 
for $\delta=0.86$ is shown.} 
\end{figure}
\noindent
stantial distance from the
TWB.
Since the rapidly
varying space and time scales have been factored out in the amplitude equations (2), 
locking tongues cannot be found therein. However on a line in the parameter space (see  Fig.\ 3a), the velocity of 
interfaces in Eqs. (2) coincides with the velocity prescribed by the locking
mechanism.
The locking mechanism arises when  
the characteristic width of the interfaces is of the same order than the
characteristic length scale of the patterns.

{\em - DPDs and their Phase Diagram. - }
Consider that a large DPD is composed of an inward-interface and an
outward-interface, that practically do not interact (see {\it e.g.}
Fig.\ 1a). 
If both interfaces exhibit no defects and are locked, 
their velocities have equal magnitude $|v_{lock}|$ but opposite signs. 
This ensures constant width and allows construction of DPDs of arbitrary size. 
Indeed, the region of existence of large DPDs starts to open 
where the locking tongues for both interface types begin to 
overlap (see Fig. 3a). 
This is the case for $a \gtrsim 5.7$.  
Above that
\begin{figure}
\label{FIG4}
\epsfxsize=70mm
\centerline{\epsffile{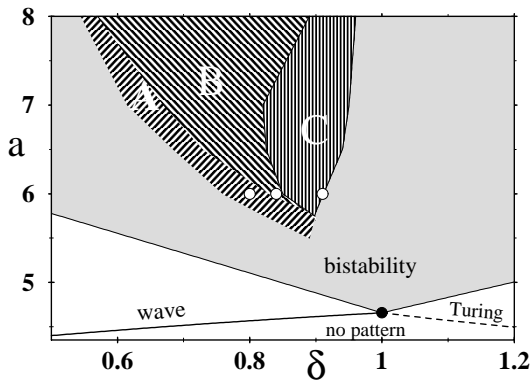}}
\caption{
The region of existence of DPDs obtained in simulations of Eq.\ (1) 
in the $a$-$\delta$ parameter space shown dashed. The light grey area
corresponds to the bistable region between the critical wave and Turing patterns
as calculated from the amplitude equations (3). In region {\bf B} DPDs  of any
size exist; the size being determined only by the initial condition. In
region {\bf C} small domains of wave patches traveling
in a Turing background are found. In region {\bf A} only
Turing-droplets are stable. The three circles correspond to the location
of simulations shown in Fig.\ 1.
}
\end{figure}
\noindent
value, DPDs spontaneously form from a variety of initial conditions.
We have determined the parameter region,
 where they propagate with constant 
width and drift speed, from extensive simulations in systems with 
sizes $L >  400$ and periodic boundary conditions. The results are shown in the phase diagram of Fig. 4. 
We can distinguish three different  subregions. 
In region B, DPDs of any size, with two
locked interfaces traveling at the same speed, are found (see Fig. 1a). 
In region A, the inward-interface is no longer locked and its speed is
smaller than $|v_{lock}|$.
Therefore large domains of Turing (wave) patterns contract (expand) 
in size until only a stable DPDs containing a single Turing cell is
left (see Fig. 1b). 
In region C, the outward-interface selects a $k_W^{sel}$ which would be 
unstable against Turing patterns in an infinite domain. 
Therefore, the wave domain forming the DPD is mostly replaced 
by a Turing pattern. 
However, small DPDs with a few wavelength of wave pattern 
are still encountered. 
At the outer boundary of region C, only DPDs with a single wave cell 
are found to be stable (see Fig. 1c). 

{\em - Conclusion - }
We found a large variety of drifting pattern domains in a
reaction-diffusion model with nonlocal coupling. 
Their ingredients include a bistability between wave and Turing
patterns near a codimension-2 point as well as absence of defects
at the interface. 
They exist as robust patterns only in a finite distance to 
the onset of pattern formation. 
Our results are not limited to the reaction-diffusion
model studied here and should carry over to other physical systems
with similar pattern forming instabilities. 
Altogether, DPDs and their constituting interfaces 
represent a generalization of simpler structures such as fronts and 
pulses in bistable reaction-diffusion systems, which do not simply 
combine two homogeneous states, but, instead, select their
constituents from whole families of possible traveling or  
stationary periodic patterns. 

\end{multicols} 

\begin{thebibliography}{99}
\bibitem{3var}
Eqs. (1) are obtained from the  three variable system
$\partial_{t}u = a u  + \beta u^2 - \alpha u^3 - bv  - g w 
+  \partial_{x}^2 u;
\partial_{t}v = c u - d v  + \delta  \partial_{x}^2 v; 
\tau_w \partial_{t}w = e u - f w  + \gamma \partial_{x}^2 w$
in the limit $\tau_w = 0$.
The parameters characterizing the nonlocal coupling in Eqs. (1)
are then found as $\sigma = \sqrt{f/\gamma}$ and $\mu = g \sqrt{e^2/\gamma f}$.
\bibitem{codim2}
We chose the parameters $b = 4, c = d = 1, \sigma = 1$ and $\mu = 2$, 
$\alpha=4/3$ and $\beta = 0$. 
%
All results have been checked for $\beta \ne 0$ and do not 
change qualitatively as long as $|\beta|$ is not too large. 
The TWB point is found at $a_{TW}=  4 \sqrt{2} -
1$ and $\delta_{TW} = 1$ 
with $k_T^c = \sqrt{2^{3/2}-1} \simeq 1.352 $,
$k_W^c =\sqrt{2^{1/2}-1} \simeq 0.644$ and $\omega_0=\sqrt{2}$.
\end{thebibliography}
\end{document}